\documentclass{aa}
\usepackage{natbib}
\usepackage{graphicx,amsmath,txfonts,url}
\usepackage{bm}
\graphicspath{{./fig/}{./png/}}

\newcommand{\EQ}{\begin{equation}}
\newcommand{\EN}{\end{equation}}
\newcommand{\EQA}{\begin{eqnarray}}
\newcommand{\ENA}{\end{eqnarray}}
\newcommand{\eq}[1]{(\ref{#1})}

\newcommand{\eqs}[2]{(\ref{#1}) and~(\ref{#2})}

\newcommand{\Sec}[1]{Section~\ref{#1}}
\newcommand{\Secs}[2]{Sections~\ref{#1} and \ref{#2}}
\newcommand{\Fig}[1]{Figure~\ref{#1}}

\newcommand{\Tab}[1]{Table~\ref{#1}}

\newcommand{\bra}[1]{\langle #1\rangle}

{}
{}
{}
\newcommand{\meanemf}{\overline{\cal E} {}}

{}
{}
\newcommand{\meanEMF}{\overline{\mbox{\boldmath ${\cal E}$}}{}}{}
{}
{}
{}
{}
{}
{}
\newcommand{\meanBB}{\overline{\mbox{\boldmath $B$}}{}}{}
{}
\newcommand{\meanFF}{\overline{\mbox{\boldmath $F$}}{}}{}
{}
{}
{}
{}
{}
{}
\newcommand{\meanJJ}{\overline{\mbox{\boldmath $J$}}{}}{}
{}
\newcommand{\meanUU}{\overline{\bm{U}}}

{}
{}
{}

\newcommand{\meanB}{\overline{B}}

\newcommand{\meanU}{\overline{U}}

\newcommand{\meanJ}{\overline{J}}

{}

{}
{}

\newcommand{\xxx}{\hat{\mbox{\boldmath $x$}} {}}
\newcommand{\yyy}{\hat{\mbox{\boldmath $y$}} {}}
\newcommand{\zzz}{\hat{\mbox{\boldmath $z$}} {}}

\newcommand{\zz}{\mbox{\boldmath $z$} {}}

\newcommand{\ww}{\mbox{\boldmath $w$} {}}

\newcommand{\kk}{\bm{k}}

\newcommand{\xx}{\bm{x}}

\newcommand{\uu}{\mbox{\boldmath $u$} {}}
\newcommand{\UU}{\mbox{\boldmath $U$} {}}

\def\bb{\bm{b}}
\newcommand{\BB}{\mbox{\boldmath $B$} {}}

\newcommand{\jj}{\mbox{\boldmath $j$} {}}
\newcommand{\JJ}{\mbox{\boldmath $J$} {}}

\newcommand{\AAA}{\mbox{\boldmath $A$} {}}

\newcommand{\ee}{\mbox{\boldmath $e$} {}}

\newcommand{\ff}{\mbox{\boldmath $f$} {}}

\newcommand{\FF}{\mbox{\boldmath $F$} {}}

\newcommand{\nab}{\mbox{\boldmath $\nabla$} {}}

\newcommand{\SSSS}{\mbox{\boldmath ${\sf S}$} {}}

\newcommand{\EMF}{\mbox{\boldmath ${\cal E}$} {}}

\newcommand{\ii}{{\rm i}}

\newcommand{\DD}{{\rm D} {}}

\def\Pm{\mbox{\rm Pr}_M}
\def\Rm{\mbox{\rm Re}_M}

\def\cs{c_{\rm s}}

\def\kf{k_{\rm f}}

\def\urms{u_{\rm rms}}

\def\nut{\nu_{\rm t}}
\def\nuT{\nu_{\rm T}}
\def\etat{\eta_{\rm t}}

\def\etaT{\eta_{\rm T}}

\newcommand{\yapj}[3]{ #1, {ApJ,} {#2}, #3}

\newcommand{\yapjl}[3]{ #1, {ApJ,} {#2}, #3}

\newcommand{\yan}[3]{ #1, {Astron.\ Nachr.,} {#2}, #3}

\newcommand{\yana}[3]{ #1, {A\&A,} {#2}, #3}

\newcommand{\ygafd}[3]{ #1, {Geophys.\ Astrophys.\ Fluid Dyn.,} {#2}, #3}

\newcommand{\ysci}[3]{ #1, {Science,} {#2}, #3}

\newcommand{\ypre}[3]{ #1, {Phys.\ Rev.\ E,} {#2}, #3}

\newcommand{\yjour}[4]{ #1, {#2}, {#3}, #4}

\newcommand{\itover}[2]{\,\hspace{.3mm}#1{\!\hspace{-.3mm}#2}}

\newcommand{\ithat}[1]{\itover{\hat}{#1}}

\newcommand{\zervec}{\bm{0}}

\newcommand{\deriv}[3]{\frac{#3\hspace*{-.06em} {#1}}{#3\hspace*{.06em} {#2}}}

\newcommand{\parder}[2]{\deriv{#1}{#2}{\partial}}

\newcommand{\parr}[2]{\frac{\partial #1}{\partial #2}}

\newcommand{\res}{\text{res}}

\newcommand {\BBM}{\overline{\BB}}

\newcommand{\aaa}{\mbox{\boldmath $a$}{}}{}

\newcommand{\zt}{k_1 z'}
\newcommand{\xk}{k_1 x}
\newcommand{\aO}{$\alpha$\,--$\,\Omega$\ }
\newcommand{\aaO}{$\alpha^2$--$\,\Omega$\ }
\newcommand{\alp}{$\alpha^2$\ }

\newcommand{\BX}{B^{\hspace*{.9pt} X}}
\newcommand{\BZ}{B^{\hspace*{.9pt} Z}}
\newcommand{\mBX}{\meanB^{\hspace*{.9pt}\raisebox{-4pt}{\scriptsize $X$}}}
\newcommand{\mBZ}{\meanB^{\hspace*{.9pt}\raisebox{-4pt}{\scriptsize $Z$}}}
\newcommand{\mmBX}{\meanB^{\hspace*{.9pt} X}}  
\newcommand{\mmBZ}{\meanB^{\hspace*{.9pt} Z}}
\newcommand{\BBZ}{\meanBB^{\hspace*{.9pt} Z}}
\newcommand{\BBX}{\meanBB^{\hspace*{.9pt} X}}
\newcommand{\JJZ}{\meanJJ^{Z}}
\newcommand{\JJX}{\meanJJ^{X}}
\newcommand{\Ba}{B^{\alpha\alpha}}
\newcommand{\Bao}{B^{\alpha\Omega}}
\newcommand{\BBa}{\meanBB^{\hspace*{.7pt}\alpha\alpha}}
\newcommand{\BBao}{\meanBB^{\hspace*{.7pt}\alpha\Omega}}

\newcommand{\BBha}{\ithat{\BB}^{\hspace*{.7pt}\alpha\alpha}}
\newcommand{\BBhao}{\ithat{\BB}^{\hspace*{.7pt}\alpha\Omega}}
\newcommand{\LAMao}{\lambda^{\alpha\Omega}}
\newcommand{\LAMaa}{\lambda^{\alpha\alpha}}

\begin{document}

\title{The fratricide of\ \aO \ dynamos by their \ \alp \ siblings}
\author{A. Hubbard\inst{1}, M. Rheinhardt\inst{1} and A. Brandenburg\inst{1,2}}
\institute{Nordita, AlbaNova University Center,
Roslagstullsbacken 23, SE-10691 Stockholm, Sweden
\and
Department of Astronomy, Stockholm University, SE-10691 Stockholm, Sweden
}

\date{\today,~ $ $Revision: 1.97 $ $}

\abstract{
Helically forced magneto-hydrodynamic shearing-sheet turbulence
can support different large-scale dynamo modes,
although the \aO mode is generally expected to dominate because it is the fastest growing.
In an \aO dynamo, most of the field amplification is produced by the shear.
As differential
rotation is an ubiquitous source of shear in astrophysics,
such dynamos are believed to be the source of most astrophysical large-scale magnetic
fields.
}{
We study the stability of oscillatory migratory \aO type dynamos
in turbulence simulations.
}{
We use shearing-sheet simulations of hydromagnetic turbulence
that is helically forced at a wavenumber that is about three times
larger than the lowest wavenumber in the domain so that both
\aO and \alp dynamo action is possible.
}{
After initial dominance and saturation, the \aO mode is found to be destroyed
by an orthogonal \alp mode sustained by the helical
turbulence alone.  We show that
there are at least two processes through which this transition can occur.
}{
The fratricide of \aO dynamos by its \alp sibling is discussed in the
context of grand minima of solar and stellar activity.
However, the genesis of \aO dynamos from an \alp dynamo has not yet been found.

}
\maketitle

\section{Introduction}

The observed existence of large-scale astrophysical magnetic fields, for example galactic or
solar fields,
is usually explained by self-excited dynamo action within electrically conducting fluids or plasmas.
However, this mechanism
of field amplification continues to be 
a matter of debate as the existing theory encounters problems when extrapolated
to the large magnetic Reynolds numbers of astrophysics.
Nonetheless,
large-scale astrophysical fields
are believed to be predominately
generated by so-called \aO dynamos,
in which
most of the field amplification occurs through the shearing of field lines by ubiquitous
differential rotation,
a process known as
the $\Omega$ effect \citep{SK69}.
For example, many models of the solar dynamo invoke the strong
shear found in the tachocline at the base of the convection zone
\citep[see, e.g.,][]{Cha10}.
Shear alone cannot drive dynamo action however, and the $\alpha$ effect,
caused by helical motions, provides
the necessary twist of the sheared  field
to complete the magnetic field amplification cycle.  In the Sun,
an $\alpha$ effect is provided via kinetic
helicity due to the interaction of stratified convection and solar rotation.

The $\alpha$ effect can drive a dynamo by itself,
being then of the so-called \alp type.
These dynamos are of great theoretical interest
due to their simplicity, but are expected to be outperformed
by \aO dynamos in the wild.
Strictly speaking, \aO dynamos
should be referred to as \aaO dynamos as
the \alp process of course continues to 
occur in reality, even in the presence of the $\Omega$ effect.
However,
in the mean-field approach one sometimes makes the so-called
``\aO'' approximation by neglecting the production of toroidal field by the
$\alpha$ effect entirely in favor of the $\Omega$ effect.
This also applies to the present paper where we consider
numerical solutions of the
original equations in three dimensions with turbulent
helical flows.  However, we will nevertheless 
refer to \aO and \aaO regimes when shear is dominant
or comparable with amplification by the helical turbulence,
respectively.

Very often, a linear stability analysis of a given setup reveals
that several different dynamo modes are expected to be excited at the same time.
While during the linear stage the relative strengths of these modes
are determined by the initial conditions,
the mode or mixture of modes of the final saturated state is decided by the quintessentially
nonlinear interactions between the modes
in their backreaction on the flow.
The naive guess that the final state should always be 
characterized by
the mode with the highest growth rate, has turned out not to be valid in general.
In \cite{Retal} it was shown for a mean-field dynamo model
with anisotropic $\alpha$
that within the appropriate parameter range both axisymmetric
equatorially anti-symmetric and non-axisymmetric equatorially symmetric
modes can be stable solutions of the non-linear system. 
For a system with differential rotation it was also  shown there 
that the stable solution can well be a mixture of axisymmetric and non-axisymmetric
modes.

In direct numerical simulations of a geodynamo model with stress-free boundary conditions,
it has been observed that again two different dynamo solutions, a dipolar and a ``hemispherical" one,
can both be stable \citep{Chetal,GB00}.
Because of the free fluid surface in that model, this might even be taken as a hint for the possibility
of non-unique stable states in stellar setups as well.
 
\cite{FRR99} have demonstrated an even more extreme case with a dynamo powered by a
forced laminar
flow.  In the course of the magnetic field growth, the
Lorentz force arranges the flow 
into a different pattern,
which is hydrodynamically stable,
but unable to
drive a dynamo.
As the dynamo dies out subsequently without a chance to recover,
it was named  ``suicidal".

Hence, the question for the character of the final, saturated stage of a dynamo cannot
reliably be answered on the basis of a linear approach and the study of the nonlinear model
might unveil very unexpected results.
Here, we will show
in a simple setup
that, while \aO dynamos do grow faster than \alp
dynamos, non-linear effects are capable of driving transitions from \aO  modes
to \alp modes.  As the two competing
dynamo modes are excited for the same
parameter set, i.e., the same
eigenvalue problem,
we refer to them as {\em fratricidal}, in reminiscence of the
aforementioned suicidal dynamos.

The two astrophysical dynamos for
which we have long time-series, the solar dynamo and that of the Earth, both exhibit large
fluctuations.  The solar dynamo in particular is known to go through prolonged quiescent
phases such as the Maunder minimum \citep{Eddy}.
A conceivable
connection with fratricidal dynamos
makes understanding how non-linear effects define
large-scale dynamo magnetic field strengths and geometries a matter of
more than intellectual curiosity.

In \Sec{MFM} we sketch the mean-field theory of \alp and \aaO dynamos.  In
\Sec{Numerics} we describe our numerical set-up and briefly discuss the test-field
method, a technique to extract the
turbulent transport coefficients
of mean-field theory from
direct numerical simulations.  In \Secs{mixed}{random} we describe
different transition types, and we conclude in \Sec{conclusions}.

\section{Mean field modeling}
\label{MFM}

In the magneto-hydrodynamic approximation,
the evolution of magnetic fields is controlled by
the induction equation
\EQ
\parr{\BB}{t} = \nab \times \left(\UU \times \BB - \eta \JJ\right), \label{induction}
\EN
where $\BB$ is the magnetic field, $\JJ=\nab \times \BB$
is the current density in units where the vacuum permeability is unity,
and $\eta$ is the microphysical resistivity.
A common
approach to \eq{induction} is mean-field theory, under which physical
quantities (upper case) are decomposed into mean (overbars) and fluctuating (lower
case) 
constituents:
\EQ
\BB = \meanBB +\bb.
\EN
The mean can be any which obeys the Reynolds averaging rules, and is frequently assumed to be
a spatial one filtering out large
length-scales
(a two-scale approach).
Here we will however use planar averaging, either over the $xy$ plane so
that
$\meanBB=\langle \BB \rangle_{xy}=\meanBB(z)\equiv \BBZ$
or over the $yz$ plane,
that is, $\meanBB=\langle \BB \rangle_{yz}=\meanBB(x)\equiv\BBX$,
where $\langle\cdot\rangle_\xi$ denotes averaging over all values of the variable $\xi$ (or volume, if not
specified). 
A mean defined by averaging over $y$ only will also be used.

Under
Reynolds averaging
Eq.\eq{induction} becomes
\begin{align}
&\parr{\meanBB}{t}=\nab \times (\meanUU \times \meanBB+\meanEMF-\eta \meanJJ), \label{inductionm} \\
&\parr{\bb}{t}=\nab \times (\meanUU \times \bb + \uu \times \meanBB + \uu \times \bb - \meanEMF -\eta \jj),
\label{inductionf}
\end{align}
where $\meanEMF \equiv \overline{\uu \times \bb}$ is the mean electromotive force (EMF) associated
with correlations of the fluctuating fields.

Symmetry considerations allow one to write the
$\meanEMF$ as a function of the mean-fields in the system.
In the case of a planar averaging scheme, the equation becomes
\EQ
\meanemf_i=\alpha_{ij}\meanB_j-\eta_{ij}\meanJ_j+\cdots, \label{alphaeta}
\EN
where $\alpha_{ij}$ and $\eta_{ij}$ are turbulent transport coefficients,
and averaged quantities
depend on one spatial coordinate only.
The traditional $\alpha$ effect is described by the symmetric part
of the tensor $\alpha_{ij}$,
and requires helicity in the flow.
The symmetric part of $\eta_{ij}$ describes turbulent dissipation, and,
in the isotropic case,
appears equivalently to the microphysical resistivity $\eta$.
It is therefore termed the
{\em turbulent resistivity}, $\etat$.
When assuming that $\meanEMF$ can be completely represented by  
 the mean magnetic field and
its
first spatial
derivatives,
the Taylor series in \eqref{alphaeta} can be truncated after the term 
$\eta_{ij}\meanJ_j$. 
A more complete formula would include higher spatial
as well as temporal derivatives.

\subsection{Mean-field dynamo action}

Let us assume a large-scale shearing flow of the simple form
\EQ
\UU_S=Sx\,\yyy. \label{eq:shear}
\EN
and velocity fluctuations 
which are isotropic,
homogeneous, and statistically stationary.
Consequently,
if
$\alpha_{ij}$ and $\eta_{ij}$ are 
assumed
to be independent of $\meanBB$ (the kinematic limit), then they reduce
to constant scalars $\alpha$ and $\etat$
\footnote{Strictly speaking, shear could
introduce
anisotropy felt by mean fields with
non-vanishing $z$-components.  Our results do not reveal any such.}.

If this system were to contain a $y$-dependent
mean field,
the shear would induce field constituents which are proportional to $x$.
We restrict ourselves here to periodic spatial dependencies and hence 
exclude such unbounded fields.  The
evolution of harmonic
mean magnetic fields
is given by the solution of the
eigenvalue problem
\EQ
\lambda\ithat\BB=
\left(\begin{array}{ccc}
-\etaT k^2 & -i\alpha k_z & 0 \\
i\alpha k_z+S & -\etaT k^2 & -i\alpha k_x \\
0 & i\alpha k_x & -\etaT k^2 \end{array}\right) \ithat\BB, \label{MFE}
\EN
where $\BBM = \ithat\BB\exp(\ii\kk\cdot\xx+\lambda t)$, $\etaT=\etat+\eta$,
and  $k^2=k_x^2+k_z^2$.
The resulting dispersion
relation
reads
\EQ
(\lambda+\etaT k^2)[(\lambda+\etaT k^2)^2-\alpha^2k^2+\ii \alpha S k_z]=0, \label{disp}
\EN
with eigenvalues
(apart from the always decaying modes with $B_y=0$)
\EQ
\lambda_{\pm}=-\etaT k^2 \pm \left(\alpha^2 k^2 -\ii \alpha S k_z\right)^{1/2}.
\label{eigen}
\EN
It can easily be seen that
there are two ``pure" modes
with particularly simple geometries:
the
 $\alpha^2$ mode with
$k_z=0$ does not depend on $S$ and has the form
\EQ
\BBha=\Ba \left( 0,\, \sin k_x x,\, \pm\cos k_x x \right), \label{Balp} 
\EN
where
the growth rate is $\LAMaa=|\alpha k_x|-\etaT k_x^2$ and
$\Ba$ is an amplitude factor.
The upper (lower) sign corresponds  to positive (negative) $\alpha k_x$.

In contrast, the \aaO
mode with $k_x=0$ does depend on $S$ and has, for $S \gg \alpha k_z$
(the \aO approximation)
the form
\begin{align}
\BBhao&=\Bao \left( \sin [k_z(z-ct)], \sqrt{2} \left| \frac{c}{\alpha}\right| \sin [ k_z (z-ct) + \phi],0\right) \label{BaO},\\
                  c&= \operatorname{sign}(\alpha S)\sqrt{|\alpha S/2k_z|}\,. \label{CaO}
\end{align}
In the above
$\Bao$ is, again, an amplitude factor,
$\phi$ represents,
for $S>0$ ($S<0$),
the
$\pm\pi/4$ ($\pm3\pi/4$) phase shift
between the $x$ and $y$ components of the mean field,
and upper (lower) signs apply for positive (negative) values of $\alpha k_z$;
see Table~3 of \cite{BS05}.
The corresponding
growth rate is
\EQ
\Re\{\LAMao\} = \sqrt{|\alpha S k_z|/2}-\etaT k_z^2\,. \label{eq:lamBZ}
\EN
For equal $|\kk|$, the \aO mode grows
faster than the \alp mode. 
\footnote{When assuming both $k_x$ and $k_z$ to be different from zero,
but keeping the \aO approximation valid and $k_z$ fixed,
the phase speed of the dynamo wave does not change while the growth rate is reduced by $\eta k_x^2$. However, 
the eigenmode has now a $z$ component $\sim  -k_x/k_z \meanB_x$.
Such modes were not observed in our simulations.} 

\begin{figure}[t!]\begin{center}
\includegraphics[width=\columnwidth]{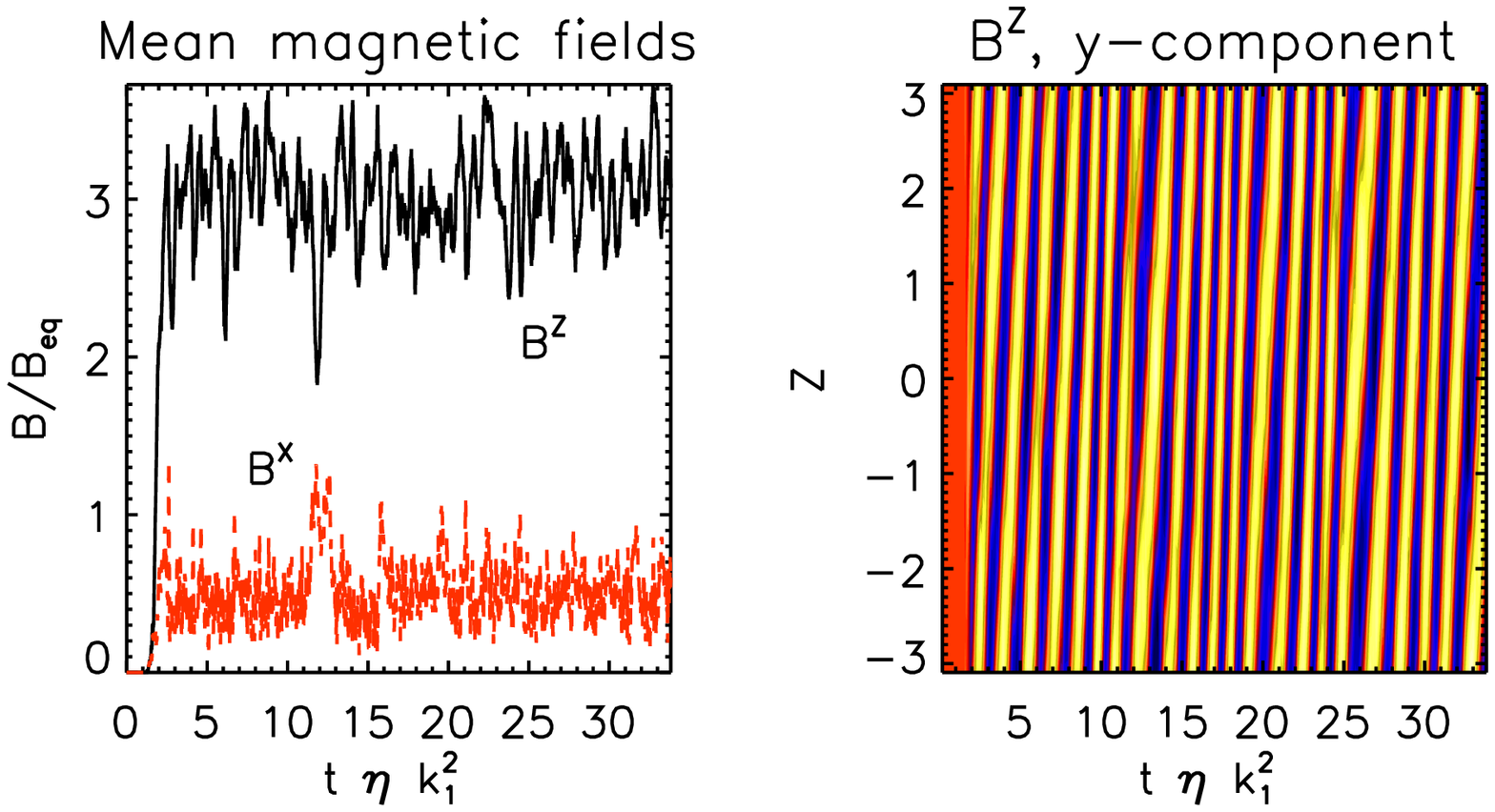}
\end{center}\caption{Time series for a
dominantly \aaO
dynamo
with $\Rm=20$, $\Pm=5$ and $\kf\approx3.1$. 
Left: rms value of $\BBZ$
defined as
$\bra{{\BBZ}^2}_z^{1/2}$,
to be associated with the \aaO mode (black/solid),
and
of $\BBX$, defined as
$\bra{{\BBX}^2}_x^{1/2}$,
to be associated with the \alp mode (red/dashed).
Right:
butterfly diagram of
$\mmBZ_y$
showing the
dynamo wave of the \aaO mode. 
\label{AObase} }
\end{figure}

A key characteristic of 
\aaO solutions is that the growth rate $\lambda$
has a non-vanishing imaginary part $k_z c$
which
results in traveling waves with phase speed $c$.  The wave
nature of \aaO solutions is a significant draw in explaining
the oscillatory solar magnetic cycle.
For  a characteristic \aaO dynamo found
numerically with a setup described below,
we show in Fig.~\ref{AObase} 
the time-series
of rms values of $\BB$
alongside the traveling wave in the $z-t$ plane (``butterfly diagram'').
This solution is
similar to those considered recently by \cite{KB09}.
There are other sources for such oscillations however.
Admittance of a spatially varying $\alpha$ enables oscillatory and hence traveling wave solutions
in pure $\alpha^2$ dynamos, see \cite{BS87}, \cite{RB87}, \cite{SG03}, \cite{Mitra10}.

The mean fields of
\alp modes are force free, while \aaO modes cause
 a potential force which
has minimal effect as long as the peak Alfv\'en speed is sub-sonic.
Within kinematics, the induction equation
allows for superimposed \alp and \aaO modes
and in \Sec{mixed} we will discuss the interactions
within such a superposition.

\section{Model and Methods}   
\label{Numerics}

\subsection{Numerical setup}

We have run simulations of helically
forced
sheared turbulence in 
homogeneous
isothermal triply (shear) periodic cubic domains with sides of length $2\pi$.
The box wavenumber, which is also the wavenumber of the observed mean fields, is therefore $k_1=1$.
Unless otherwise specified, our simulations have $64^3$ grid points.
For the shear flow we have taken the one defined by \eqref{eq:shear}.
We solve the non-dimensionalized system
\begin{align}
&\parr{\AAA}{t}=\UU \times \BB+\eta \nab^2 \AAA \label{dAdt} \\
&\frac{\DD\UU}{\DD t} =  -\cs^{2}\nab\ln{\rho} + \frac{1}{\rho}\JJ\times\BB
 +\FF_{\rm visc}+ \ff,  \label{dUdt}\\
&\frac{\DD\ln\rho}{\DD t}=  -\nab\cdot\UU, \label{dRdt}
\end{align}
where $\cs=1$ is the isothermal sound speed, $\rho$ the density,
$\FF_{\rm visc} = \rho^{-1}\nab\cdot(2\rho\nu\SSSS)$ the viscous force,
${\sf S}_{ij}=\frac{1}{2}(U_{i,j}+U_{j,i})-\frac{1}{3}\delta_{ij}\nab\cdot\UU$
is the rate of strain tensor, $\nu$ is the kinematic viscosity and
$\ff$ the forcing term.
We use the {\sc Pencil Code}\footnote{http://pencil-code.googlecode.com},
which
employs
sixth-order explicit finite differences in space and a third
order accurate
time stepping method.
While our code
allows full compressibility
 the simulations
are only weakly compressible (small Mach number). 
As in earlier work \citep{Brandenburg2001ApJ}, in each time step the forcing function 
is a snapshot of a circularly polarized plane wave.   All these waves
have the same handedness, but
their direction and phase change randomly from
one time step to the next.  This forcing provides
kinetic helicity.
The wavevectors are taken from the set of vectors that satisfy periodicity 
and whose
moduli are adequately
close to the target forcing wavenumber $\kf$. 

The magnetic vector potential is initialized with a weak Gaussian random field,
the initial velocity is given by $\UU=\UU_S$ and the initial density is uniform.
 In \Tab{params} we have collected the control parameters and some key derived
quantities of the model.
Two parameters of note are the magnetic Reynolds and Prandtl numbers,
\EQ
\Rm=\urms/\eta\kf,\quad
\Pm=\nu/\eta.
\EN
To characterize the turbulence, we provide values of $\alpha$ and $\etat$ which
characterize the corresponding
tensors as described in \Sec{MFM}. 
These were determined using the
test-field method with test-field wavevector
$\kk=\ithat{\xx}$ or $\kk=\ithat{\zz}$.   

\begin{table}[b!]
\caption{Control and derived parameters}
\label{params}
\centerline{\begin{tabular}{l@{\hspace{4mm}}l@{\hspace{4mm}}l}
\hline \hline
$\nu$ & Control par. & Microphysical viscosity \\
$\eta$ & Control par. &  Microphysical resistivity \\
$ S $ & Control par. & Shear ($\UU_S=Sx \ \yyy$)  \\
$f_{\rm rms}$ &  Control par. & Forcing amplitude \\
$\kf$ &  Control par. & Forcing wavenumber (generally $\kf \approx3.1$) \\[1mm]
\hline \\[-2.5mm]
$\Pm$ & $\nu/\eta$  & magnetic Prandtl number\\
$\urms$ & $\bra{\uu^2}^{1/2}$ &  RMS turbulent velocity \\
$\Rm$ & $\urms/\eta \kf$ &  Magnetic Reynolds number \\
$k_1$ & $k_1=1$ & Wavenumber of mean fields (box wavenumber) \\
$t_\res$ & $1/\eta k_1^2$ & Resistive time (mean fields) \\
$t_{\mathrm{turb}}$ & $1/\urms\kf$ & Turbulent time \\   
\hline \hline
\end{tabular}}
\end{table}

For our purposes, we require
the helical turbulence to be
strong enough
that the $\alpha^2$ dynamo can 
safely be excited. 
For this we guaranteed that in all our simulations,
$\Rm$ is above the critical value (of the order of unity) for \alp dynamos
in the corresponding \emph{shearless} setup.
Further, some of the transitions we will study require long simulation times
due to their rarity,
which constrains us to modest numerical resolutions.  This in turn prevents
our (explicit) numerical resistivity from being small, so the
turbulent velocities must be reasonably large for
the stated super-critical values of $\Rm$.
Choosing furthermore subsonic shear speeds,
we are restricted to a modest region of parameter space.
In light of these limitations we operate mostly in a $\Pm>1$ regime.

\subsection{Test-field method}
\label{testfieldxz}

A fundamental difficulty in extracting
the tensors $\alpha_{ij}$ and $\eta_{ij}$ from a numerical simulation of
\eqref{dAdt}--\eqref{dRdt} is that \eqref{alphaeta} is under-determined.
Turbulent transport depends on the velocity field,
so ``daughter'' simulations of the induction equation,
whose velocity fields are continuously copied from the main run,
share the same tensors $\alpha_{ij}$ and $\eta_{ij}$.
It is therefore possible to lift the degeneracy
by running an adequate number of daughter simulations
with suitably chosen       
``test'' mean fields.    
We employ this \emph{test-field method} (TFM);
for an in depth overview see
\cite{Sch05,Sch07} and \cite{BRS08,BRRK08}.
Recently the original method has been extended to
systems with rapidly evolving
mean-fields,
requiring
a more complicated \emph{ansatz} than Eq.~{alphaeta}
\citep{Memory} and to
the situation with magnetic background turbulence
\citep{RB10}. 

In addition to calculating planar-averaged turbulent tensors as described in the references above,
we will be interested in tensors that depend both on $x$ and $z$ (that is, are $y$-averages).
For this,
we generalize \eqref{alphaeta} to
\EQ
\meanemf_i=\alpha_{ij}\meanB_j+\beta_{ijk}\parder{\meanB_{j}}{x_k}+\cdots. \label{alphabeta}  
\EN
There are $27$ tensor components (as $\partial_y \meanBB=\zervec$), so nine test-fields are required,
which we choose
to be
of the form
\EQ
   \BB^{pq} = B^{\rm T} f_q(x,z) \delta_{ip} \ithat{\ee}_i, \quad p=1,2,3, \quad q \in \{ \rm cc, sc, cs\},
\EN
where $f_q(x,z)$ is defined, according to the choice of $q$,
to be one of the following functions:
\[
\cos k_1 x \cos  k_1 z ,\quad \sin  k_1 x \cos  k_1 z,\quad \cos  k_1 x \sin  k_1 z ,
\]
and $B^{\rm T}$ is, as standard for test-field methods, an arbitrary amplitude factor.
Although the wavenumber of the test fields is usually treated as a varying parameter
we need here to consider only the single value $k_1$ because the fastest growing 
and also the saturated dynamos
in the simulations are dominated by
this wavenumber, the smallest possible in our periodic setup.
As is often the case in applications of the test-field method, we will occasionally be faced with unstable solutions 
of the test problems.
We treat that difficulty by periodically resetting the
test solutions \citep[see][]{Imposed}.
Since it takes a finite time for the test solutions to reach their stationary values, 
and as this time is frequently
close to the required reset time,
only limited windows
in the time series
of the data are valid.
 
\section{Dynamical interactions of \alp and \aaO modes}
\label{mixed}

\begin{figure}[t!]\begin{center}
\includegraphics[width=\columnwidth]{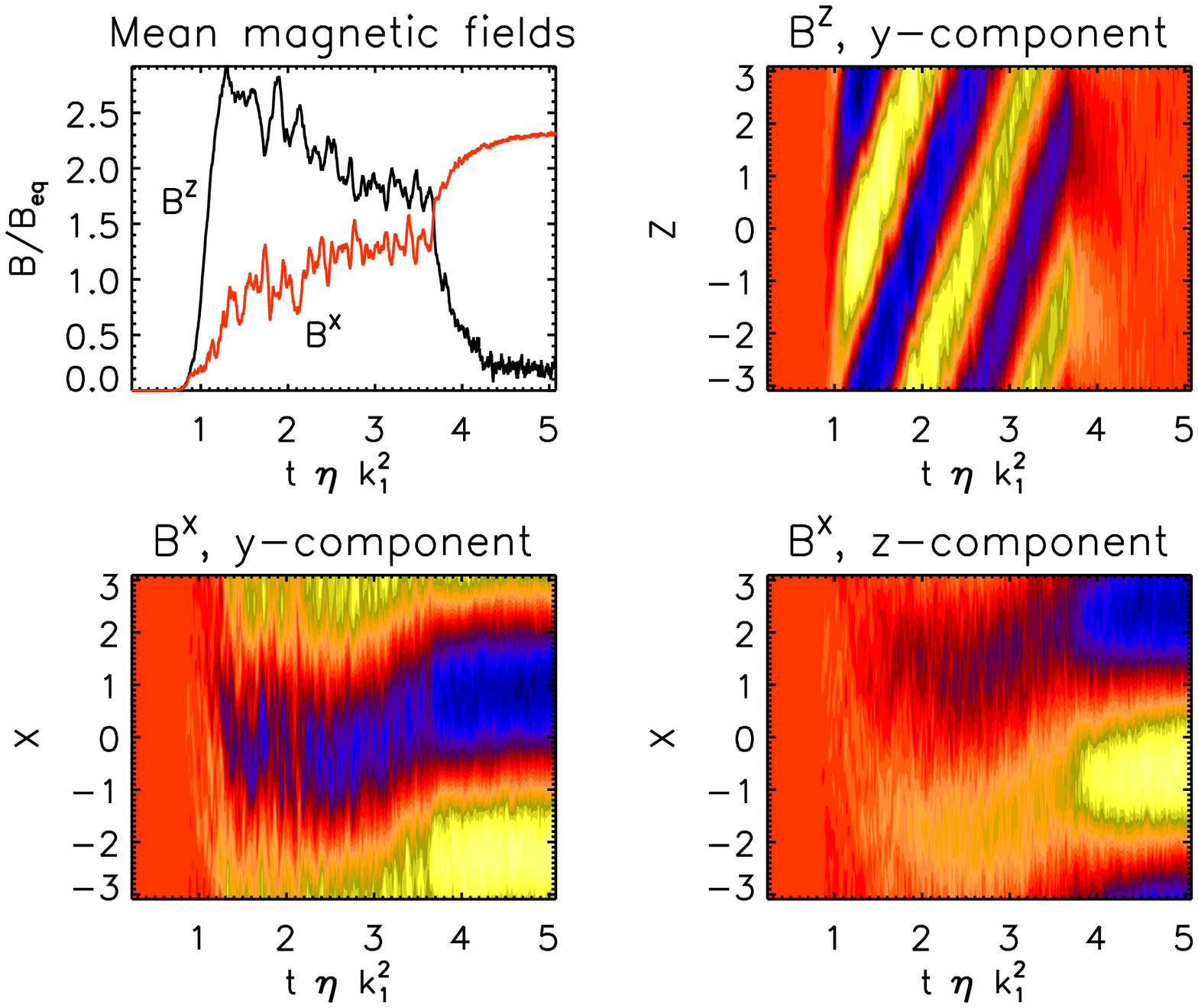}
\includegraphics[width=\columnwidth]{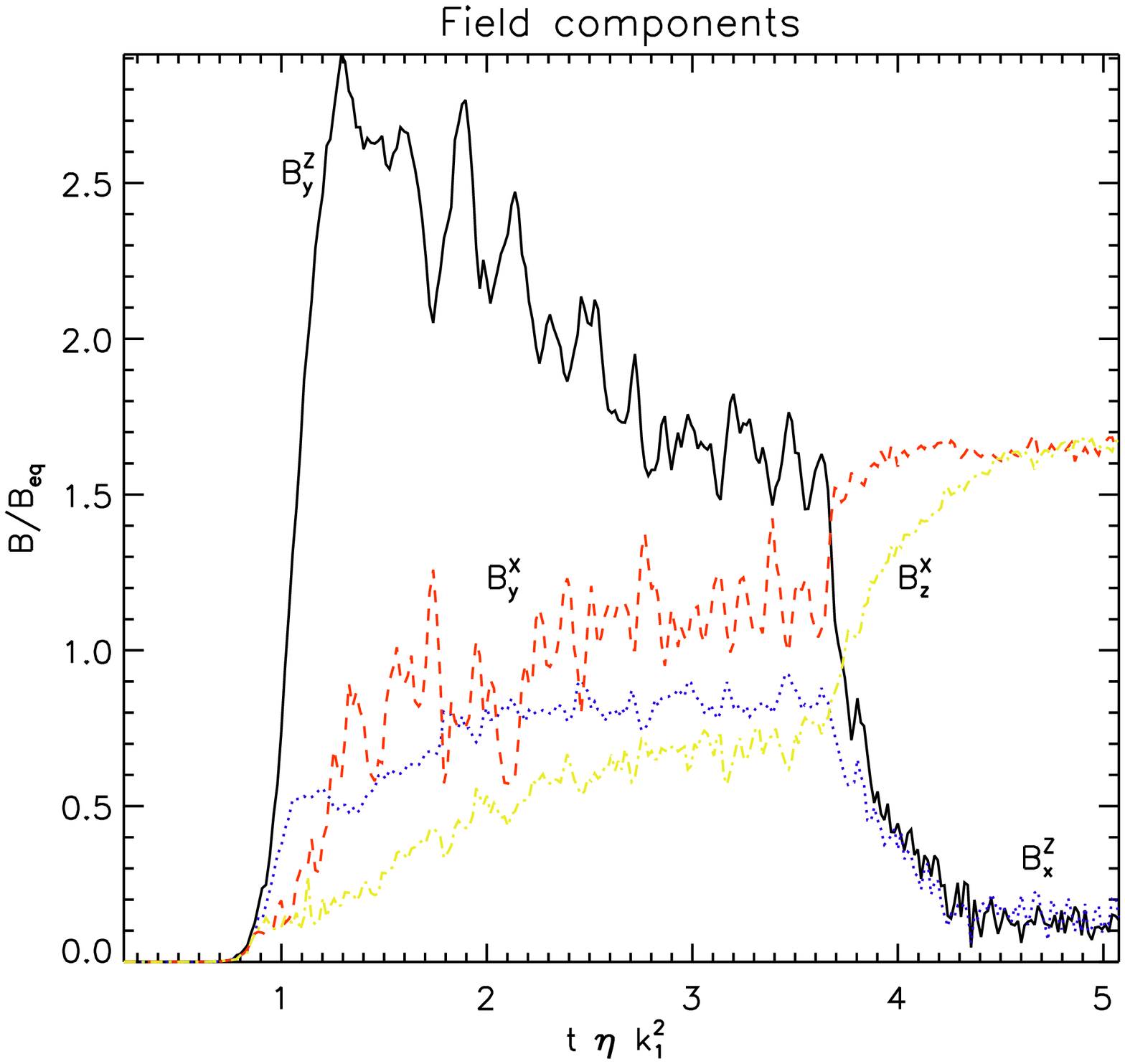}
\end{center}
\vspace{2mm}
\caption{Time series for Run A. 
Upper row: same quantities as in Fig.\ \ref{AObase}.
Middle row:
$\mmBX_y$ and $\mmBX_z$, to be associated with the \alp mode.
Note that the \aaO and \alp modes
coexist
during the transition.
Lower panel:
rms values of the components of  $\BBX$ and $\BBZ$.
\label{RunA}}
\end{figure}

\begin{figure}[t!]\begin{center}
\includegraphics[width=\columnwidth]{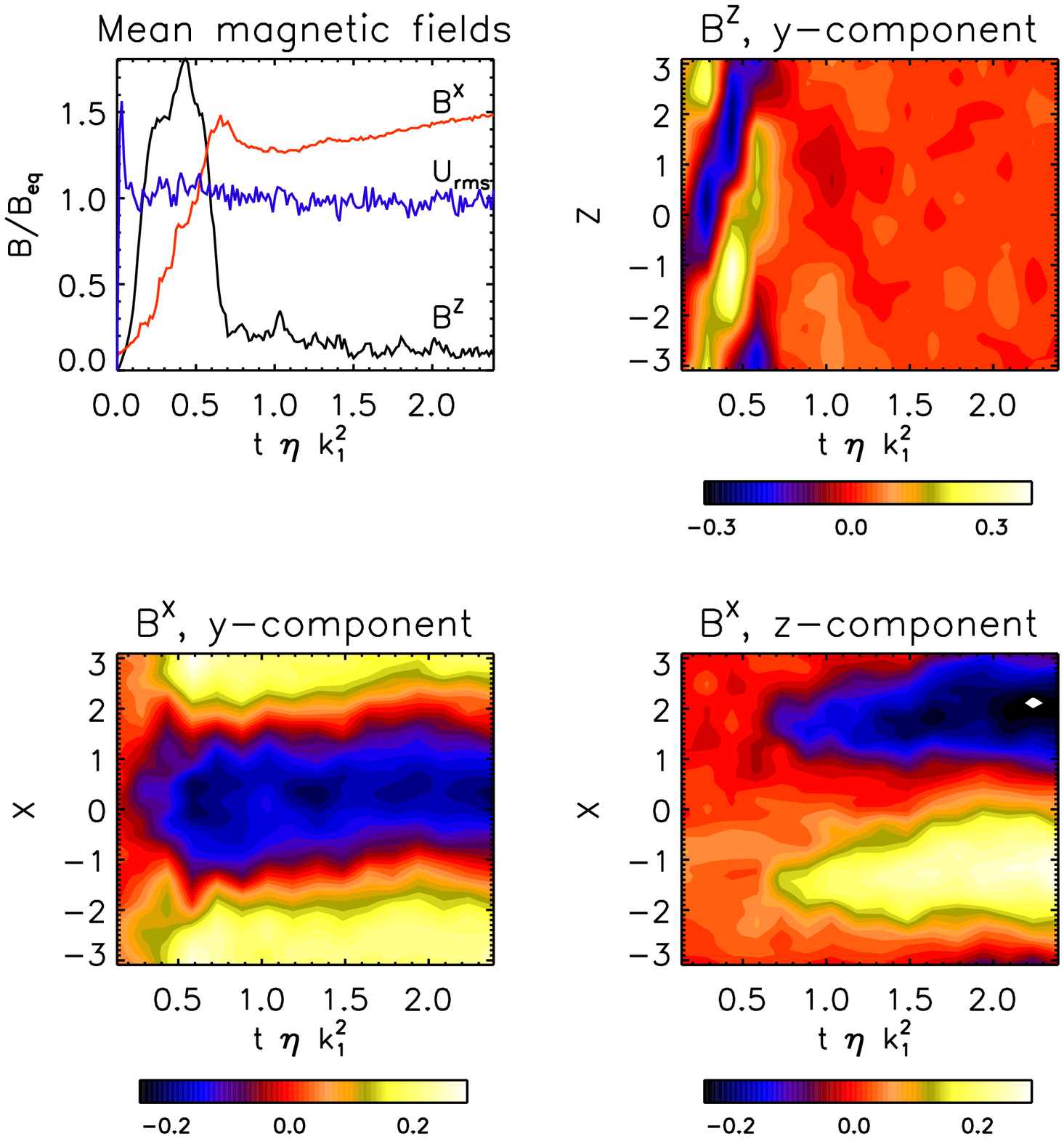}
\end{center}

\vspace{5mm}
\caption{Time series for Run E that never entered a
quasi-stationary
 \aaO regime. 
Top left: rms values of $\BBZ$ to be associated with the \aaO mode (black) and
of $\BBX$ to be associated with the \alp mode (red).
Note the  considerably faster growth of the \aaO mode during the kinematic phase.                                                                                                                
Top right: Butterfly diagram of 
$\mmBZ_y$, showing the traveling
dynamo wave
during the kinematic phase only, but later merely fluctuations.     
Bottom: components of $\BBX$.                                                                            
\label{noAO}}
\end{figure}

Here we report on the results of our simulations
a first set of which is characterized in \Tab{runs1}.
In \Fig{RunA} we show
time series for Run A,
which saw a transition from
a $z$-varying \aaO dynamo
($\BBZ$)
 to an $x$-varying \alp dynamo
 ($\BBX$).
As is made clear in the bottom panel, there was a prolonged period where
the two modes were coexisting
while their relative strengths were changing
monotonically.
However, note that $\BX_y$ is stronger than $\BX_z$, that is, 
the \alp field is distorted
during the transition.
Run A was repeated
16 times with the same parameters, but 
different random seeds, and all
these runs
exhibited similar behavior.
Likewise we performed runs where
both the value of $\eta$ and the
numerical resolution (cf. Runs B-D, I,J)
were varied.
As these additional runs
also showed
the same transition
pattern, we conclude that
it is deterministic for this level of shear and forcing.
More, we conclude that for these
cases the \aaO mode
is unstable to the growth of an \alp mode due
to non-linear effects.
Runs with the dynamical 
parameters ($S$, $\urms$) of \Tab{runs1}
inevitably generate \alp fields 
from \aaO fields
after modest times, runs with significantly different parameters will usually 
(for most of the
random seeds)  
exit the kinematic regime into an \aaO mode, and stay in that mode for
a prolonged time with no sign of an \alp field.
Nonetheless even such simulations can occasionally fail to fully enter in the \aaO regime, instead exiting the
kinematic regime into an \alp mode, as shown in \Fig{noAO}.

\begin{table}[b!]
\caption{Run parameters \label{runs1}}
\centerline{\begin{tabular}{lcccccccc}
\hline \hline
Run & Res. & $-S$  & $\urms$ & $-\alpha^\dagger$ &
$\etat^\dagger$ & $\Rm$ & $\Pm$ &
 $\tau^\ddagger$ \\
\hline
Run A & $64^3$   & $0.05$ & $0.11$ & $0.04$ & $0.025$ & $37$ & $5$ & $2$--$3$\\
Run B & $64^3$   & $0.05$ & $0.17$ & $0.04$ & $0.03$ & $26$ & $2.5$ & $2$ \\
Run C & $\!\!128^3\!\!$ & $0.05$ & $0.14$ & $0.04$ & $0.027$ & $44$ & $3$ & $4$ \\
Run D & $\!\!128^3\!\!$ & $0.05$ & $0.14$ & $0.04$ & $0.027$ & $90$ & $6$ & $1.5$ \\
Run E & $64^3$   & $0.02$ & $0.13$ & $0.04$ & $0.023$ & $90$ & $10$ & N/A \\
Run I   & $64^3$ & $0.05$ & $0.15$ & $0.04$ & $0.036$ & $49$ & $1$ & $3$ \\
Run J  & $64^3$  & $0.05$ & $0.19$ & $0.04$ & $0.035$ &  $31$ & $0.5$ & $1.5$ \\
\hline \hline
\end{tabular}}
\tablefoot{$\dagger$ Time-averaged values
determined through the TFM using 
harmonic
test fields with $\kk=\ithat{\xx}$ or $\kk=\ithat{\zz}$.
The results are identical due to
homogeneity of the time-averaged turbulent velocity. 
$\ddagger$ $\tau=t_{\mathrm{dur}}/t_{\res}$ is the duration of the
transition of the type discussed in \Sec{mixed}; counting
from multiple
random seeds for Run A.}
\end{table}

\subsection{Mean-field approach}
\label{MFAppr}
Clearly, 
the transition from an \aaO mode to an \alp one
must be a consequence of the back-reaction of $\meanBB$ onto the flow.
Within the mean-field picture, there are two channels available
for it:
(i) the back-reaction onto the fluctuating flow, usually described as  a
dependence of $\alpha_{ij}$ (more seldom $\eta_{ij}$) on the
mean field and
(ii) the back-reaction onto to the mean flow by the mean Lorentz force,
which might again be decomposed into a part resulting from the fluctuating field,
$\overline{ \jj \times \bb}$, and one resulting from the
mean field, $\meanJJ \times \meanBB$.
Here, we will deal with a flow generated by the latter force
that straddles the distinction of means and fluctuations:
it survives under $y$-averaging,
but vanishes under the $xy$ and $yz$ averaging that reveals the \aaO and \alp dynamos
respectively.
For simplicity we consider magnetic
field configurations that would result from a superposition of
linear modes of the \aO and \alp dynamos, given in equations \eqs{BaO}{Balp}
respectively.
Such a situation will inevitably occur during the kinematic growth phase
if both
dynamos are supercritical, but
is only relevant for analyzing the back-reaction onto the flow
if it at least to some extent
continues into the non-linear regime.
Our
analysis is linear in nature, so while it provides a qualitative framework for understanding
the transition process,
it is
surely not quantitatively accurate.

In order to be able to consider both $\BBX$ and $\BBZ$ as mean fields under
one and the same averaging, we have now to resort to $y$ averaging.
Moreover, for the sake of clarity we will occasionally subject the 
resulting $x$ and $z$ dependent mean fields further to spectral
filtering with respect to these coordinates. That is, we will consider
only their first harmonics $\sim \mathrm{e}^{\ii k_1 (x +z)}$ as mean fields.

Let us represent the mean field $\langle\BB\rangle_y$ as 
superposition of 
a $\BBX$ resembling the ($x$ varying) \alp mode $\BBa$ (Eq.~\eqref{Balp})
and a $\BBZ$ resembling the ($z$ varying) \aO mode $\BBao$  (Eq.~\eqref{BaO}):
\begin{alignat}{2}
\BBZ&=\BZ\begin{pmatrix}\sin(\zt)\\ G\sin(\zt+\phi) \\ 0 \end{pmatrix},\;\;
&\JJZ&=k_1\BZ \begin{pmatrix} -G \cos(\zt+\phi) \\ \cos(\zt) \\ 0\end{pmatrix}, \nonumber \\
\BBX&=\BX\begin{pmatrix} 0 \\ H\sin \xk \\ \cos \xk \end{pmatrix},\quad
&\JJX&=k_1\BX \begin{pmatrix} 0 \\ \sin \xk \\ H\cos \xk \end{pmatrix},
\label{mixedmodes}
\end{alignat}
with $z' \equiv z-ct$
 recalling that $c$ is the speed of the dynamo wave (Eq.~\eqref{CaO}).
In the above,
 $\pi/4 \le \phi \le 3\pi/4$ 
and $G,H, k_1 > 0$ are appropriate for $\alpha>0$,
The parameters $G$ and $H$ capture the
difference in the
strengths of the $y$ 
and $z$ components (\alp) or
the $x$ or $y$ components (\aaO), respectively.
We expect $G>1$ as
shear amplifies the $y$
component
of an
\aaO
mode
well above its $x$ component. 
The inclusion of the parameter $H$, which is unity for pure \alp
modes
will be justified below,
but can already
 be seen in the different strengths
shown in \Fig{RunA}, lower panel.

\begin{figure}[t!]\begin{center}
\includegraphics[width=\columnwidth]{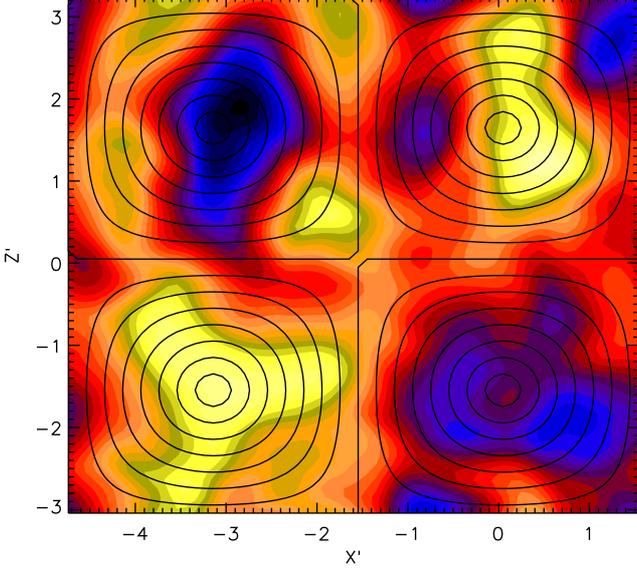}
\end{center}\caption{
$\langle U_y\rangle_y$ for Run A, 
taken at early time ($t=1.45 t_\res$)
when $\BBX$ is still of only modest strength.
Plotting area is shifted in $x$ and $z$ 
to make the quadrupolar
geometry clear.
Overplotted contours: quadrupolar constituent
$\sim \cos k_1 x' \cos k_1 z'$    
\label{quad1}}
\end{figure}

The mean Lorentz force
 $\meanJJ \times \meanBB$ for the
superimposed fields can be written as
\begin{align}
\meanFF_L &= \bra{\FF_L}_y = \\&\phantom{+}k_1\BX\BZ \cos \xk [G \cos (\zt+\phi) +H \sin (\zt)] \yyy
             +\nab \Phi.   \nonumber
\end{align}
As the Mach numbers were found to be small throughout,
we assume incompressibility and hence
drop the potential component $\nab \Phi$.
Further, we assume that $\meanFF_L$ and the
mean velocity driven by it are simply linked by a coefficient
$K \approx 1/\nuT k_1^2$, where the total viscosity $\nuT$ is the sum of the molecular
$\nu$, and
the turbulent viscosity $\nut$.
Thus we can approximate the mean
velocity due to the interaction of the
superimposed mean fields as
\EQ
\meanUU_L
=U_L\cos\xk\left[G\cos(\zt+\phi)+H\sin(\zt)\right]\yyy, \label{Uquad}
\EN
where $U_L=Kk_1\BX\BZ$. 
Clearly, this flow, having merely a $y$ component, shows quadrupolar geometry in the $x-z$ plane as
$\meanU_{L,\,y}$ can be rewritten
in the form 
$U'_L \cos\xk \cos(\zt+\phi')$ with  a new amplitude and phase, $U'_L$ and $\phi'$.

The simulations
show indeed a dominant part of that shape in the Lorentz-force  generated mean flow as can be seen
from  \Fig{quad1}. There the y-averaged $U_y$ is shown together with its Fourier constituent
$\sim \mathrm{e}^{\ii k_1(x+z)}$.
The latter contains approximately $10\%$ of the energy in this component, or
$U'_L=
\meanU_{y,\mathrm{rms}}/3$, 
indicating that the assumptions made in deriving \eqref{Uquad} are 
reasonably well justified in a non-linear system.

Upon interaction with a $\BBX$ or a $\BBZ$ of the form \eqref{mixedmodes},
the mean flow $\meanUU_L$ in \eqref{Uquad}
generates an $\meanemf_x(z)$ and $\meanemf_z(x)$, respectively.

\subsection{Dominating \aaO mode}    

If $\BZ \gg \BX$, then
$\BBX$ can be treated as a perturbation, and we can
drop higher order terms in $\BX$.  
Accordingly, the $z$-averaged EMF due to the flow $\meanUU_L$ is
\EQ
\EMF^X=\bra{\meanUU_L \times \BBZ}_z=
\frac{Kk_1}{2}\BX {\BZ}^2 (G\sin\phi-H)  \cos \xk\;\zzz.
\EN
The curl of this EMF is
\EQ
\nab \times \EMF^X= \BX I \sin(\xk)\yyy , \quad
I \equiv \frac{K k_1^2 {\BZ}^2}{2}\left(G\sin\phi-H\right). \label{defI}
\EN
If $G\sin \phi>H$, then $I>0$ and
for $H>0$
this EMF reinforces 
$\mBX_y=\BX H\sin \xk$. 
Thus we see that
the inclusion of the parameter $H$
in the ansatz for $\BBX$, Eq.~\eqref{mixedmodes}, 
was justified as
$\mBX_y$
receives enhanced forcing
in comparison to $\mBX_z$.

\subsection{Dominating \alp mode}   

\label{domalp}
If $\BX \gg \BZ$ then we can
in turn
treat $\BBZ$ as a perturbation.  Further, as the system is
dominated by
the \alp
mode, we will have $H \sim 1$.  In this case we
find
\begin{eqnarray}
\EMF^Z=&&\bra{\meanUU_L \times \BBX}_x \\
=&&\frac{Kk_1}{2}\BZ {\BX}^2\left[G\cos(\zt+\phi)+H\sin(\zt)\right]\xxx, \nonumber
\end{eqnarray}
and
\EQ
\nab \times \EMF^Z=\BZ\,\frac{K k_1^2 }{2}{\BX}^2\left[H\cos(\zt)-G\sin(\zt+\phi)\right]\yyy. 
\EN
We can write
\begin{align}
&\hspace*{-3mm}H\cos(\zt)-G\sin(\zt+\phi)=   \label{AlpMF} \\
&\hspace*{-3mm}\left[(H\cos\phi-G)\sin(\zt+\phi)\right]_1+\left[H\cos\phi\cos(\zt+\phi)\right]_2. \nonumber
\end{align}

\vspace{2mm}\noindent
If $H\cos\phi-G<0$, as expected since $H\sim 1$,
$G>1$,
term $[]_1$ in \eqref{AlpMF} will act to damp 
$\mBZ_y$, that is,
the perturbative \aaO wave.  Further, term $[]_2$ is opposite in sign to the time-derivative of
such a
wave, so it slows or reverses the direction of wave-propagation.

\subsection{Mean-Field Evolution}
\label{MFI}

Here we assume
again domination of the \aaO mode, that is,
$\BZ \gg \BX$.
With Eqs.~\eqref{MFE} and \eqref{defI}
the eigenvalue problem for the modified \alp field $\BBX$ is then
(adopting $k_x$ = $k_1$, $k_z=0$)
\EQ
\lambda^X\BBX=
\left(\begin{array}{ccc}
-\eta_T k_1^2 & 0 & 0 \\
S & -\eta_T k_1^2 & -i(\alpha k_1+I) \\
0 & i\alpha k_1 & -\eta_T k_1^2 \end{array}\right) \BBX,
\EN
with eigenvalues
\EQ
\lambda^X=-\eta k_1^2 \pm \sqrt{\alpha k_1(\alpha k_1+I)}.
\EN
Making the approximation $I \gg \alpha k_1$, similar to the $\alpha$-$\Omega$
approximation
$S \gg \alpha k_1$, we find
\EQ
\lambda^X=-\eta k_1^2 \pm \sqrt{\alpha I k_1}.
\EN
The above should be compared with
the growth rate of the \aO  
dynamo,
$\LAMao$
from \eqref{eq:lamBZ}
which is not touched by the occurrence of $I$.
The \aO
dynamo saturates when $\alpha$ has been quenched such that the product $\alpha S$
settles at 
the marginal value
$|\alpha S|=2 \eta_T^2 |k_1|^3$.
If the parameter $I$
becomes comparable with the shear, i.e.,
$ I  \sim S$,
then
$\BBX$
might grow even when
the \aaO field is saturated, i.e. $\lambda^X>\Re (\lambda^{\alpha\Omega})=0$.
In other terms,
the \aaO mode is unstable to the growth of a \emph{fratricidal}
\alp field, so the transition will take a well defined time from the
onset of the non-linear stage, determined by $\lambda^X$.

We test this theory for Run A
at the time of Fig.~\ref{quad1}, $t = 1.45t_\res$,
extracting $G$ and $H$ from the relative strengths of the
$x$ and $y$  or $y$ and $z$ components
of the averaged fields
$\BBZ$ or $\BBX$, respectively, after a projection onto the first harmonics; see Eq.~\eqref{mixedmodes}. 
The parameter $I$ is calculated from the magnetic and velocity fields using
\EQ
I=\frac{k_1 B^Z U_L}{2B^X}\left(G\sin \phi-H\right),
\EN
with 
$U_L=U'_L/\sqrt{G^2+H^2-2GH \sin\phi}$,
where $U'_L$ is the amplitude of the quadrupolar
constituent
of the velocity field seen in Fig.~\ref{quad1}.
We find
$U'_L\simeq 0.07$,
$H \simeq 2.9$, $G \simeq 4.9$, $I \simeq 0.09$, and confirm that $\phi \simeq \pi/4$.
As
$I>S=0.05$, the growth of the $x$-varying mode even when the \aO mode is saturated
is not surprising.
Repeating this run (keeping the
control parameters
fixed) 
16 times with different random seeds changed the
occurrence
time of the transition by only one resistive
time, suggesting that the transition is an essentially deterministic process.

We have never seen a reverse transition
from the \alp state  back to the \aaO state.
This may be understood
in terms of interacting
modes, with the \aaO mode being suppressed
once the \alp mode is dominating; see Sec.~\ref{domalp}.

\section{Random transitions}
\label{random}

\begin{figure}[t!]\begin{center}
\includegraphics[width=\columnwidth]{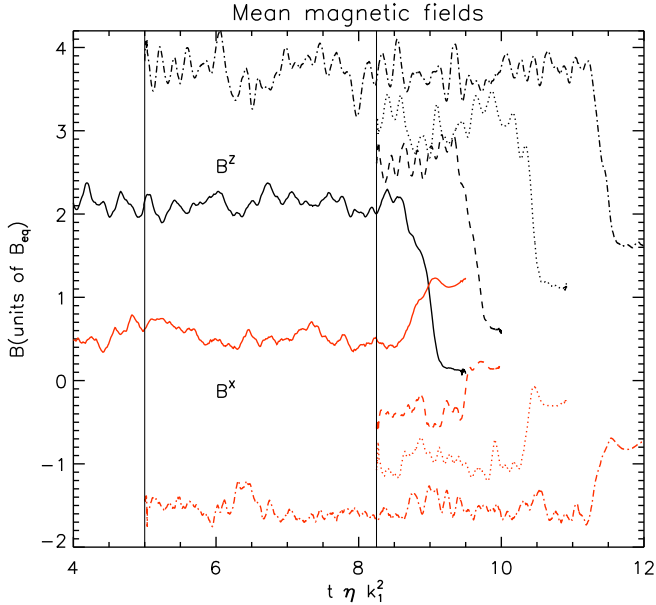}
\end{center}\caption{Time series for Run F (solid lines), with
rms values of $\BBZ$ (black) and $\BBX$ (red). 
Broken lines: restarts from the main run
with new random seeds, vertically offset for visibility.
All the runs end up with the same energies in $\BBX$ and $\BBZ$.     
Vertical lines: restart times.
\label{TS:random}}
\end{figure}
\begin{table}[b!]
\caption{Run parameters}
\label{runs2}  
\centerline{\begin{tabular}{lcccccccc}
\hline \hline
Run & $\!\!$Res.$\!\!$ & $-S$  & $\urms$ & $-\alpha^{\dagger}$ & $\etat^{\dagger}$ & $\Rm$
& $\Pm$ & $\tau^{\ddagger}$ \\
\hline
Run F & $64^3$ & $0.2$ & $0.14$ &  $0.02$ & $0.1$ & $90$ & $10$ & $5, 9^{\ddagger}$\\
Run G & $64^3$ & $0.2$ & $0.099$&  $0.01$ & $0.16$ & $63$ & $10$ & $5,20^{\ddagger}$ \\     
Run H & $64^3$ & $0.1$ & $0.085$ & $0.017$& $0.037$ & $27$ & $5$  & $25$\\  
\hline \hline
\end{tabular}}
\tablefoot{$\dagger$ see \Tab{runs1}.
$\ddagger$ $\tau=t_{\mathrm{trans}}/t_{\res}$ is the time when transition occurred;
for F and G 
over multiple realizations with differing random seeds.
See Fig.~\ref{scatter}.}
\end{table}

\begin{figure}[t!]\begin{center}
\includegraphics[width=\columnwidth]{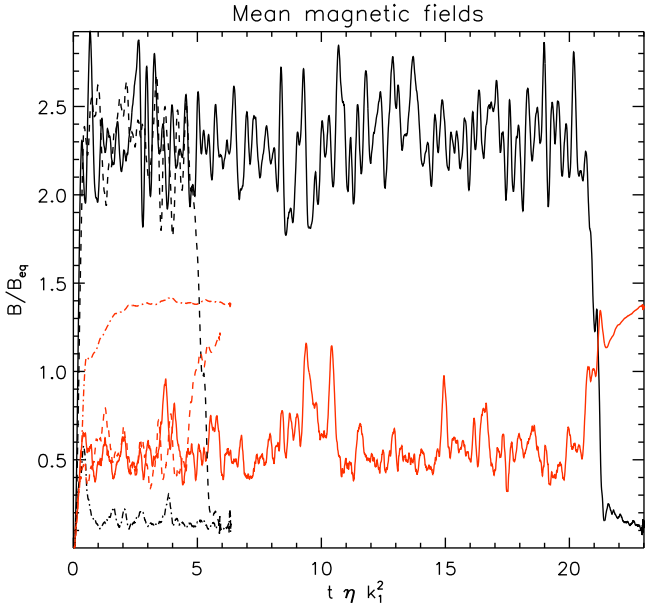}
\end{center}\caption{Time series of Run G (solid) along with
a sibling run (dashed) with different seeds, 
showing significant differences in the transition start time.
Dash-dotted:
a run which never entered the \aaO regime.
\label{tsmult}}
\end{figure}

\begin{figure}[t!]\begin{center}
\includegraphics[width=\columnwidth]{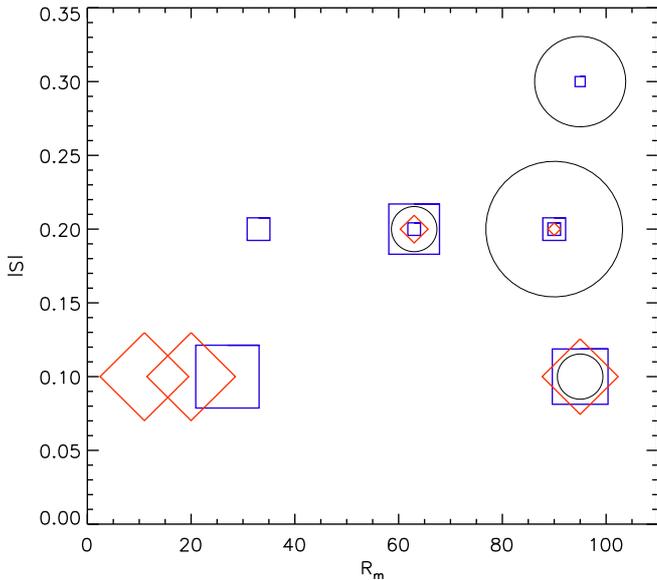}
\end{center}\caption{
Synopsis of runs which did not
exhibit the instability discussed in \Sec{mixed}.   
Runs at the same position differ only in random seeds.
Circles/black: a significant \aaO mode was never developed
(cf. Fig.~\ref{noAO}), size indicates the corresponding
number of runs ($1$, $2$ or $3$).
Square/blue: a transition occurred, size represents
the time until transition ($4$ to $25 t_{\res}$).
Diamond/red: the \aaO stage was entered, but no
transition occurred.  Size represents the time
span of simulation ($5$ to $35 t_{\res}$).
\label{scatter}}  
\end{figure}

Not all transitions fit the above deterministic picture of   
interacting \aaO and \alp
modes.
In Fig.\ref{TS:random} we present a set of time series of
the rms values of $\BBZ$ and $\BBX$, all related to Run F of \Tab{runs2}.
Secondary runs were performed by branching off from the original simulation either at $t=5t_\res$,
when the \aaO mode is well established and stationary,
or at $t=8.25 t_\res$, immediately before the transition to the \alp mode is launched.
The only difference between all these runs is
in the random seed, which is
used
by the forcing algorithm. 
In all, the time until the transition starts varies 
by $\approx2.5 t_\res$,
and many more
turbulent turnover times ($\Rm\kf^2/k_1^2 \simeq 800$ turbulent turnover times per resistive time).
The time elapsed during a transition is
always
of the order of
$t_\res/2$,
unlike
$3t_\res$
for the process seen in Fig.~\ref{RunA}. 
Thus it is suggestive to assume 
that there might be a very slow,
still
essentially  deterministic process, 
preparing the transition, which is
likely resistive in nature as that is the longest obvious ``native'' timescale of the system.  Slow resistive
effects are known to exist in dynamos, for example
the slow resistive growth of
 \alp dynamos in periodic systems.
However, transitions can indeed occur at very different times
including the extreme case in which
a run never develops a quasi-stationary \aaO mode,
but instead enters the \alp state almost immediately after the end of the kinematic phase, see
 Fig.~\ref{tsmult}
(run G of \Tab{runs2}).  We believe therefore that
under certain circumstances 
the transition process is not a deterministic one, in that it is impossible to predict
or at least estimate
the time until the transition. 
\Fig{scatter} is a synopsis
of simulations that belong to that type, hence do not show
the instability discussed in \Sec{mixed}.  
Note that,  while corresponding setups without shear are
known to enable \alp modes for the entire parameter range,
the \alp mode is possibly
sub-critical for $\Rm=10, S=-0.1$.

This is different from the interacting mode picture of Sec.~\ref{mixed}
 in several interesting ways.  Firstly, the \aaO mode
is here at least meta-stable against growth of the \alp mode, as evinced by
its prolonged life-time (hundreds of turbulent times)
and the small magnitude of $\BBX$, 
which further is not dominated by a \alp mode.
A reasonable working hypothesis for the cases of Sec.~\ref{mixed} is then that, there,
the \alp mode is the only stable solution and, as soon as the nonlinear stage has been entered,
it starts to devour the \aaO one, settling after a time which is related to basic parameters of the 
system and hence not random.
In contrast, for the cases considered here, we conclude that both the \alp {\em and} the \aaO
solutions are indeed stable (not only metastable) and the latter has a well extended basin of entrainment.
Due to its higher growth rate the system settles first in the \aaO mode and suppresses
the \alp mode efficiently.
A transition to the latter can only occur if a random fluctuation in the forcing is strong enough to push
the system over the separatrix into the basin of entrainment of the \alp mode.
This can happen after a rather long time only or immediately after the end of the linear stage
which has both been observed.

Given that the examples for the first scenario (\Tab{runs1}) differ from those for the second (\Tab{runs2}) mainly
in their lower rate of shear, our conclusion seems reasonable as stronger shear should result in a clearer
preference of the \aaO mode as the \alp mode does not feel the shear.
Or, in other terms, from a certain shear rate $S$ on, the \aaO mode should acquire a basin
of entrainment with a finite ``volume" that grows with $S$.
If this picture is true, transitions in the two scenarios should have clearly different characteristics,
and indeed,  the transition in Fig.~\ref{TS:random} is markedly faster than that seen in Fig.~\ref{RunA}.

As in the transitions discussed in Sec.~\ref{mixed}, we have not here seen the
\alp mode transit back into the \aaO mode.
Some attempts were made to
provoke this reverse transition by perturbing the \alp state with a (sufficiently strong) \aaO mode.
While in some runs it indeed took over,   
velocities were attained
for which the numerics are unreliable, and often proved
numerically unstable, making the results inconclusive.
However, such a behavior is not entirely surprising as the \aaO saturation process
can anyway be somewhat wild, cf. Fig.~\ref{noAO}.

\begin{figure}[t!]\begin{center}
\includegraphics[width=\columnwidth]{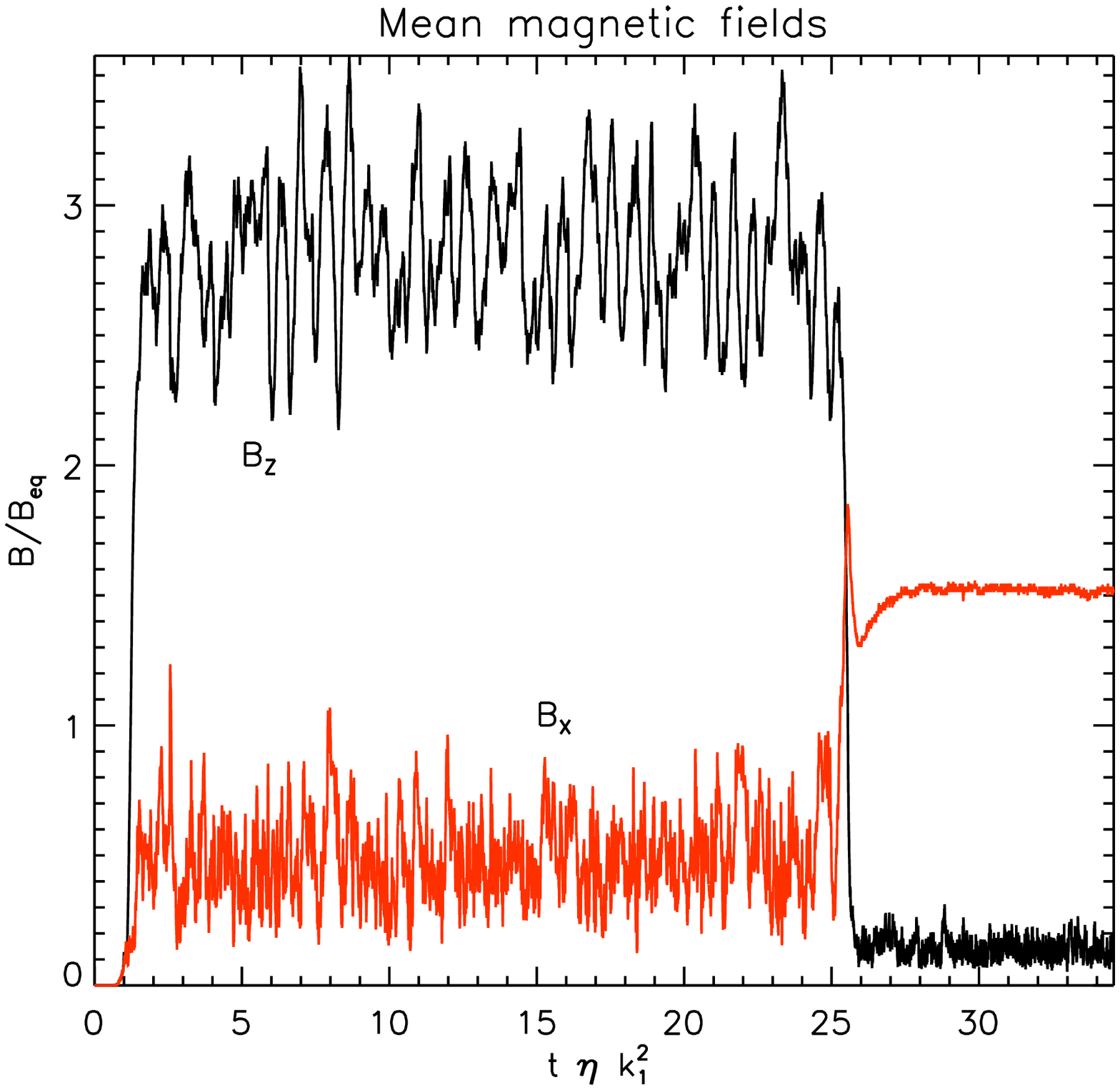}\\[10mm]
\includegraphics[width=\columnwidth]{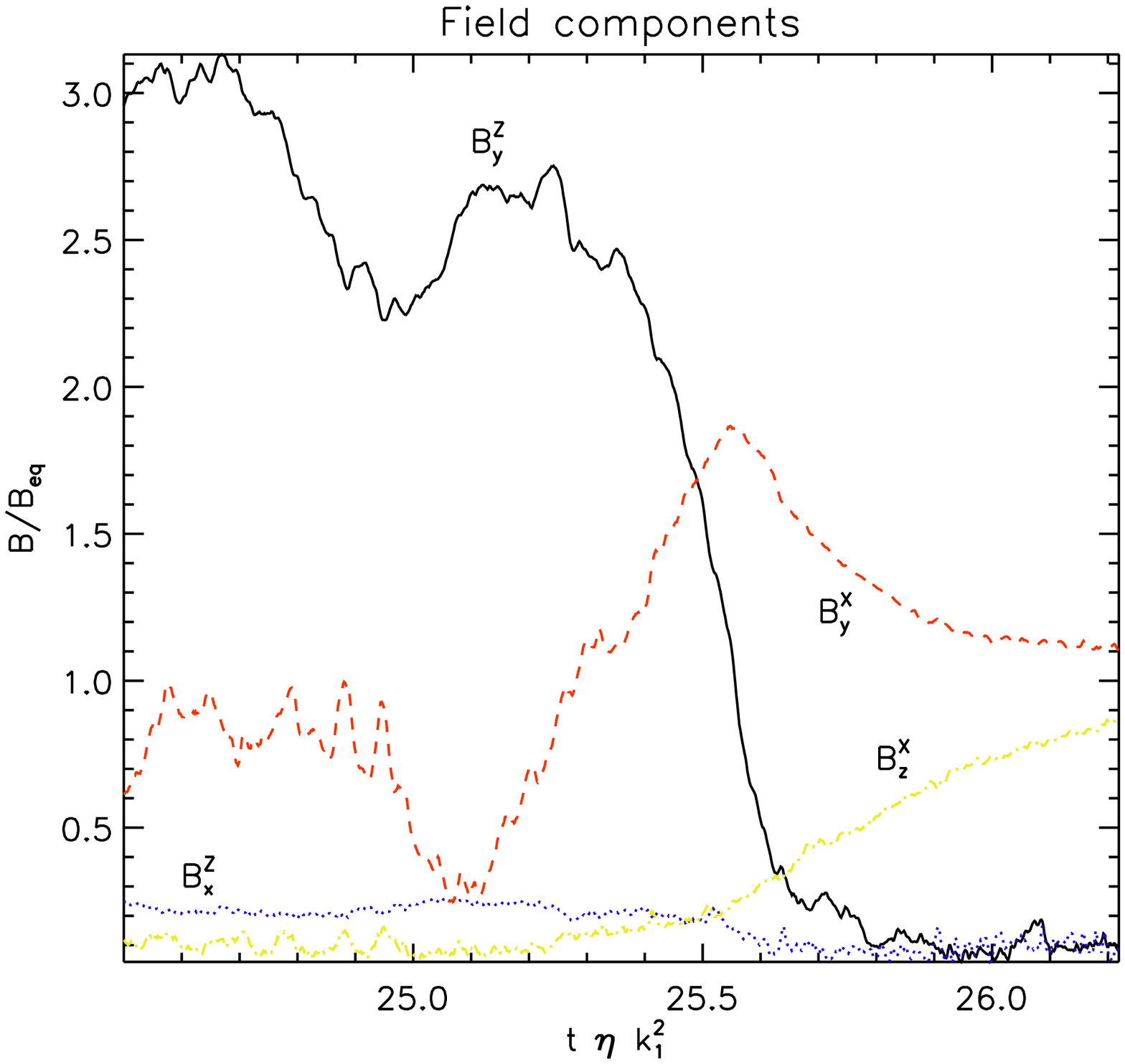}
\end{center}\caption{
Time series of Run H. Upper panel: rms values of $\BBX$ and $\BBZ$.
Note the long time before the
transition starts in comparison to Run F (see Fig.~\ref{TS:random})
and the dramatic difference in the fluctuation 
levels before
and after the transition.
Lower panel: rms values of the components.
Note the strong difference between $\mmBZ_y$ and $\mmBZ_x$, expected for an \aaO field.
More significantly, notice that $\mmBX_y$ develops before $\mmBX_z$.
\label{TS:fourth_x}}
\end{figure}

The absence of {\em spontaneous} reverse transitions appears plausible insofar  the time variability of the \alp mode
is much smaller than that of the \aaO mode, which can clearly be seen in Fig.~\ref{TS:fourth_x} for Run H. 
That is, events capable of pushing the system over the separatrix are simply much rarer. Significantly longer
integration times are likely needed for their eventual detection, but it is also conceivable that the triggering event 
never shows up.

\subsection{Large scale patterns}

\begin{figure}[t!]\begin{center}
\includegraphics[width=\columnwidth]{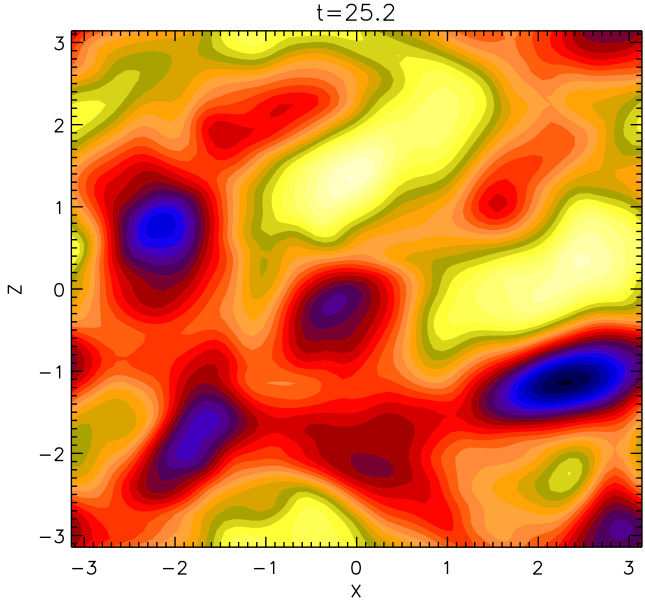}
\end{center}\caption{
$\bra{U_y}_y$ taken
during the
transition of Run H
shown in Fig.~\ref{TS:fourth_x} ($t=25.2 t_\res$).
Note the lack of a
quadrupolar geometry.
\label{UY:fourth_nex2var2}}
\end{figure}

\begin{figure}[t!]\begin{center}
\includegraphics[width=\columnwidth]{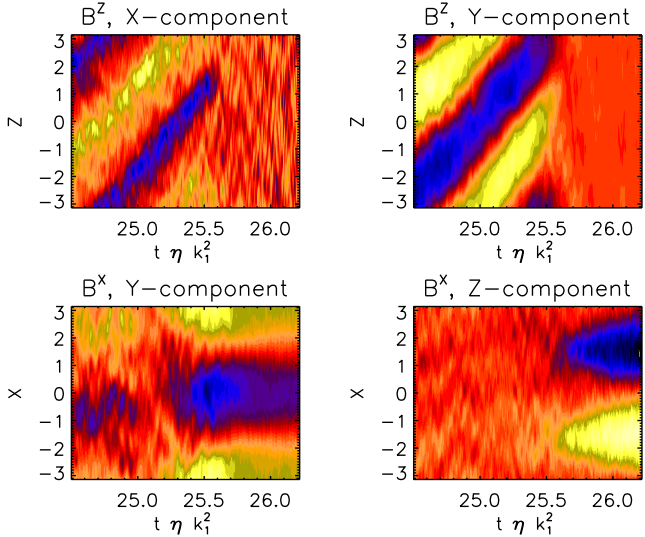}
\end{center}\caption{Butterfly diagrams for 
Run H, (see Fig.~\ref{TS:fourth_x}).
Note that
$\mmBX_y$ develops before $\mmBX_z$, i.e.,
$\meanBB$
does not
transit
from an \aaO field straight to an \alp
one.
\label{fourth_butterfly}}
\end{figure}

Run H will be examined
here in more detail.
Curiously, 
$\bra{U_y}_y$ taken just during the transition as
 shown in Fig.~\ref{UY:fourth_nex2var2}
does not show the quadrupolar pattern of
Fig.~\ref{quad1}. 
It is therefore not surprising that the butterfly diagrams in Fig.~\ref{fourth_butterfly} do not show a direct transition 
from the \aaO 
to the \alp dynamo,  as 
$\mBX_z$
develops significantly
later than
$\mBX_y$.  This is
clearly visible
in Fig.~\ref{TS:fourth_x}, lower panel.
As consideration of 
the mean flow due to the Lorentz force of the mean 
field alone is obviously not fruitful in explaining this
 transition, 
 we recall that the back-reaction of the mean field onto the turbulence opens another  channel of nonlinear interaction.

\begin{figure}[t!]\begin{center}
\includegraphics[width=\columnwidth]{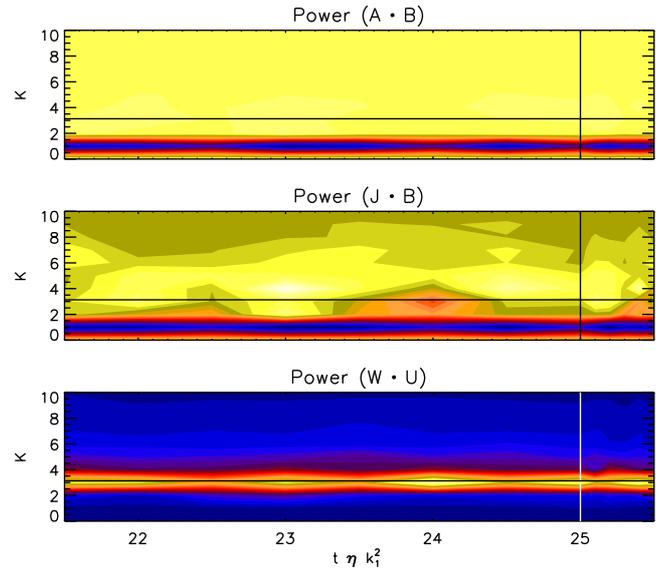}
\end{center}\caption{Time series of the
helicity power spectra
for Run H.  Horizontal line: 
forcing wavenumber
$\kf\approx3.1$.
Vertical line: border between
low-resolution observations (every $0.5 t_\res$)
for $t<25 t_\res$
and higher-resolution observations (every $0.1t_\res$ )
for $t>25 t_\res$.
Possible features in
the latter range
are
likely
due to
the increased temporal resolution.
\label{spec}}
\end{figure}

According to elementary mean-field
dynamo theory, the $\alpha$
effect is caused 
by the helicity in the flow:
 $\alpha \sim \bra{\ww \cdot \uu}$, where
 $\ww \equiv \nab \times \uu$
 is the
 fluctuating
 vorticity.  Further, the back-reaction of
the mean
field on the turbulence, which saturates the dynamo, is
assumed to be
captured by the current helicity $\bra{\jj \cdot \bb}$.
It is often related to the magnetic helicity
$\bra{\aaa \cdot \bb}$
and thought to reduce the original $\alpha$ by producing a {\em magnetic}
contribution of opposite sign. 
In Fig.~\ref{spec} we present time-series of the power spectra of these helicity correlators across the transition.
We see no clear signal
around the transition event.

\subsection{Mean-field modeling with $y$ averaging}

To examine the problem more closely, we
recall Eq.~\ref{alphabeta} 
for when the mean is defined by a $y$ average:  
\EQ
\EMF_{i}(x,z)=\alpha_{ij}(x,z) \meanBB_j(x,z)+\beta_{ijk}(x,z)\meanBB_{j,k}(x,z). \label{alpbeta}
\EN
It is clear that
the Fourier constituents
of $\alpha_{ij}$ and $\beta_{ijk}$
with
wavenumber
$k_1$ in both $x$ and $z$ (the quadrupolar constituents) can create an
emf
$\EMF^X$
out of a field
$\BBZ$,
both
with
the same
wavenumber:
\begin{align}
  \meanemf^X_i= \left\langle\alpha^{11}_{i1}\mBZ_x + \alpha^{11}_{i2} \mBZ_y  +  \beta^{11}_{i13} 
  \parder{\mBZ_x}{z} + \beta^{11}_{i23} \parder{\mBZ_y}{z}\right\rangle_z ,
\end{align}
where the superscripts indicate the coefficients to be the Fourier constituents $\sim \mathrm{e}^{\ii k_1(x+y)}$
and $\mBZ_z$ is assumed to vanish. 
Note that each of them is actually given by four values, e.g.,
the two amplitudes and phases
in:
\EQ
 \alpha^{11}_{ij} = \alpha_{ij}^\mathrm{c}\cos(k_1 x + \phi_{ij}^{\mathrm{c}})
 \cos k_1 z +  \alpha_{ij}^\mathrm{s}\cos(k_1 x + \phi_{ij}^{\mathrm{s}})\sin k_1 z.  \label{coeffs}
\EN

\begin{figure}[t!]\begin{center}
\vspace{9mm}
\includegraphics[width=\columnwidth]{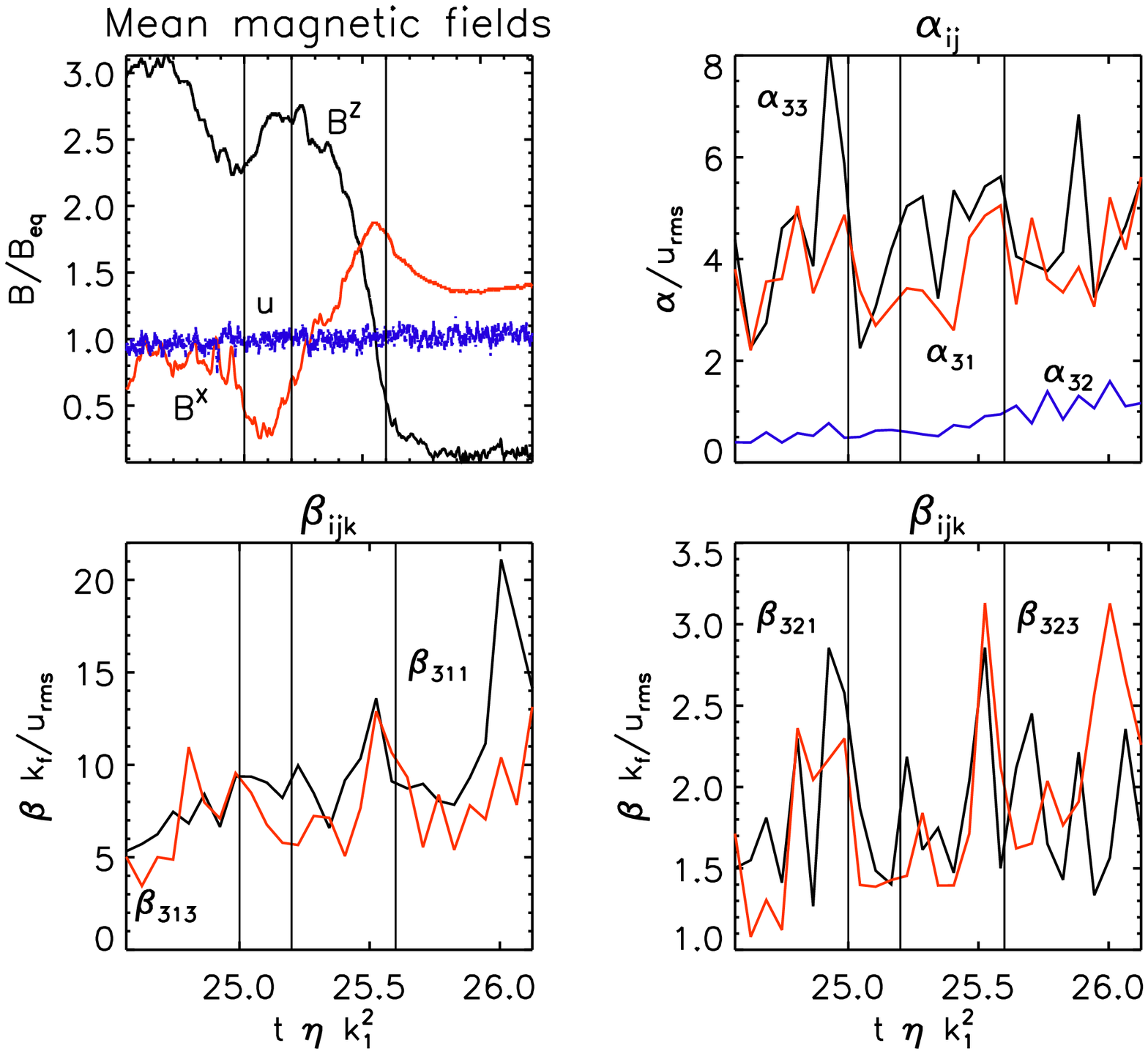}
\end{center}\caption{
Run H.
Upper left panel: rms values of $\BBX$, $\BBZ$ and $\uu$
($\urms = \bra{\uu^2}^{1/2}$),
cf. Fig.~\ref{TS:fourth_x}.    
Remaining
panels:
Selected
quadrupolar moments of $\alpha_{ij}$ and $\beta_{ijk}$
determined
by the test-field method
and
given by $({\alpha_{ij}^\mathrm{c}}^{\!2}+{\alpha_{ij}^\mathrm{s}}^{\!2})^{1/2}$
(see Eq.~\eqref{coeffs}),
likewise for $\beta_{ijk}$.
Normalization is by 
the temporally averaged $\urms$, 
as $\urms$ undergoes a slow, steady drift over time.
Vertical lines
mark the times of
events
visible
in the first panel.
\label{alpbetafig}}
\end{figure}

The coefficients relevant for the generation of $\mBX_y$ (from $\meanemf^X_z$ only) are 
$\alpha^{11}_{31},\;\alpha^{11}_{32},\;\beta^{11}_{313}$ and $\beta^{11}_{323}$.
We have used the test-field method (see Sec.~\ref{testfieldxz})
to find
them
and present the results
in Fig.~\ref{alpbetafig}.   
They turn out to be surprisingly large,
when compared to the rms velocity (e.g., 
$({\alpha^{c}_{31}}^{\!\!2} +  {\alpha^{s}_{31}}^{\!\!2})^{1/2}
\gtrsim 4 \urms$)
and some may show a trend across the transition from the \aO
to the \alp mode. This overall trend
is hypothesized to be due to the increase in $\urms$ that accompanies the
transition from a stronger \aO field to a weaker \alp field
with less potential to inhibit the flow.
It is interesting that with the exception of $\alpha_{33}^{11}$, 
the large transport coefficients are all those which generate an $\meanEMF$ out of the $\ithat{x}$-directed field, i.e.
out of a field that feels the effect of shear.  We speculate that these coefficients, with, themselves, explicit
$x$-dependence, feel the shear quite strongly.

\section{Discussion and conclusions}
\label{conclusions}

We have demonstrated that,
while \aaO modes are
kinematically preferred to \alp modes
in homogeneous systems that support both, 
the \alp mode acts 
in a fratricidal manner
against the former
after the nonlinear stage has been reached.
This transition
can occur in at least two different fashions.
Further, we have not observed the reverse
process.
One of the
two transition
processes, based on
superposed
\aaO and \alp modes, operates
in a basically deterministic fashion                  
through a large-scale velocity pattern generated
by the interaction of the modes.
In contrast, we interpret the mechanism of the second process,
which may start only many resistive times past
the saturation of the  \aaO dynamo,
by assuming that both the \aaO and the \alp modes are stable solutions of the nonlinear system.
Transitions occur if due to the random forcing a sufficiently strong perturbation builds up
which tosses the system out of the basin of entrainment of the \aaO mode into that of the
\alp mode.
This hypothesis is bolstered by both the random timing of these transitions and by
the large time-variability seen in the amplitude of the \aaO field.
A return seems to be much less likely as the level of fluctuations of the \alp mode is,
by contrast, greatly reduced.

These results fit with earlier work studying dynamos whose non-linear nature is fundamentally different
from their linear one
 \citep[e.g.,][]{FRR99}. 
While our simulations are limited to Cartesian, cubic, 
shearing-periodic domains, they are particularly exciting given that the only dynamo which has been
observed over a long baseline and which could be 
either \aO or $\mbox{\alp}\!\!$,
the solar dynamo, indeed
shows differing modes of operation
(regular cycles vs. deep minima).
The results are also disturbing in that we have evidence for non-deterministic,
rare (as they 
occur in scales of multiple resistive times or
hundreds of turbulent turnovers) mode changes that show no evidence
for a return.  Given that
the \alp mode in our simulations seems much calmer than the \aaO mode,
a rare random excursion in the field geometry is likely to be the initiating agent of the transition. 
While a bifurcation between different stable modes has long been an acknowledged
possibility for dynamos \citep{Betal89,Jen91},
a rare, stochastic, possibly uni-directional
transition
is perhaps the most troublesome form of such bifurcations    
except for the ultimate self-extinction.
 
The \aO dynamo is believed to be common and important for systems like the Sun or accretion disks,
which all have long life-times compared to turbulent
turnover times.
It is then a daunting possibility that we
could be forced to stretch our simulations over
very long temporal base-lines to find the actual long-lasting
field configuration.    
More positively, our
result, while in a different geometry, increases the importance of recent work on non-oscillatory \aO   
and oscillatory \alp modes in spherical shells for the solar dynamo
\citep{Mitra10,Sch11}.

\begin{acknowledgements}
We acknowledge the allocation of computing resources provided by the
Swedish National Allocations Committee at the Center for
Parallel Computers at the Royal Institute of Technology in
Stockholm and the National Supercomputer Centers in Link\"oping.
This work was supported in part by
the European Research Council under the AstroDyn Research Project No.\ 227952
and the Swedish Research Council Grant No.\ 621-2007-4064.
\end{acknowledgements}


\end{document}